\documentclass[journal,twoside,]{IEEEtran}
\ifCLASSINFOpdf
\else
\fi
\usepackage{xfrac}
\usepackage{mathtools,amsmath}
\usepackage{multirow}
\usepackage{color}
\usepackage{amssymb}
\usepackage{cite}
\DeclareMathOperator*{\argmin}{argmin}
\usepackage{hyperref}
\usepackage{lipsum}
\usepackage{algpseudocode,algorithm}
\DeclareMathAlphabet{\mathpzc}{OT1}{pzc}{m}{it}
\usepackage[cal=dutchcal]{mathalfa}
\ifCLASSOPTIONcompsoc
\usepackage[caption=false,font=normalsize,labelfon
t=sf,textfont=sf]{subfig}
\else
\usepackage[caption=false,font=footnotesize]{subfi
g}
\fi

 \usepackage{etoolbox}

\usepackage{stfloats}
 \makeatletter
\patchcmd{\@maketitle}
  {\addvspace{0.5\baselineskip}\egroup}
  {\addvspace{-.8\baselineskip}\egroup}
  {}
  {}
\makeatother

\begin{document}
\bstctlcite{IEEEexample:BSTcontrol}

% paper title
% Titles are generally capitalized except for words such as a, an, and, as,
% at, but, by, for, in, nor, of, on, or, the, to and up, which are usually
% not capitalized unless they are the first or last word of the title.
% Linebreaks \\ can be used within to get better formatting as desired.
% Do not put math or special symbols in the title.

\title{\makebox[\linewidth]{\parbox{\dimexpr\textwidth+1.5cm\relax}{\centering   DeepMuD: Multi-user Detection for Uplink Grant-Free NOMA IoT Networks via Deep Learning}}}

% \title{DeepMuD: Multi-user Detection for Uplink Grant-Free NOMA IoT Networks via Deep Learning}
%
%
% author names and IEEE memberships
% note positions of commas and nonbreaking spaces ( ~ ) LaTeX will not break
% a structure at a ~ so this keeps an author's name from being broken across
% two lines.
% use \thanks{} to gain access to the first footnote area
% a separate \thanks must be used for each paragraph as LaTeX2e's \thanks
% was not built to handle multiple paragraphs
%
% \author{Ferdi Kara, ~\IEEEmembership{Member,~IEEE,}
%         Hakan Kaya %~\IEEEmembership{Fellow,~OSA,}
%         % <-this % stops a space
% \thanks{This work is supported by Zonguldak Bulent Ecevit University with grant no: 2019-75737790-01}% <-this % stops a space
% \thanks{The authors are with Wireless Communication Technologies Laboratory (WCTLab) and Department of Electrical-Electronics Engineering at Zonguldak Bulent Ecevit University, Zonguldak, Turkey,67100, e-mail: \{f.kara,hakan.kaya\}@beun.edu.tr}}

\author{Ahmet Emir, Ferdi Kara,~\IEEEmembership{Member,~IEEE,} Hakan Kaya, Halim Yanikomeroglu,~\IEEEmembership{Fellow,~IEEE.} 
 \thanks{A. Emir, F. Kara and H. Kaya are %with Wireless Communication Technologies Laboratory (WCTLab)
with the Electrical-Electronics Engineering, Zonguldak Bulent Ecevit University, Zonguldak, Turkey,  e-mail: \{ahmet.emir,f.kara,hakan.kaya\}@beun.edu.tr.}
 \thanks{F. Kara and H. Yanikomeroglu are with the Department of Systems and Computer Engineering, Carleton University, Ottawa, K1S 5B6, ON, Canada, e-mail:halim@sce.carleton.ca.}}\vspace{-7.5\baselineskip}

\vspace{-3.5\baselineskip}
\maketitle
\vspace{-7.5\baselineskip}
% As a general rule, do not put math, special symbols or citations
% in the abstract or keywords.
\begin{abstract}
In this letter, we propose a deep learning-aided multi-user detection (DeepMuD) in uplink non-orthogonal multiple access (NOMA) to empower the massive machine-type communication where an offline-trained Long Short-Term Memory (LSTM)-based network is used for multi-user detection. In the proposed DeepMuD, a perfect channel state information (CSI) is also not required since it is able to perform a joint channel estimation and multi-user detection with the pilot responses, where the pilot-to-frame ratio is very low. The proposed DeepMuD improves the error performance of the uplink NOMA significantly and outperforms the conventional detectors (even with perfect CSI). Moreover, this gain becomes superb with the increase in the number of Internet of Things (IoT) devices. Furthermore, the proposed DeepMuD has a flexible detection and regardless of the number of IoT devices, the multi-user detection can be performed. Thus, an arbitrary number of IoT devices can be served without a signaling overhead, which enables the grant-free communication.  
 %up to $\sim10-20$ dB
\end{abstract}
% Note that keywords are not normally used for peerreview papers.
\begin{IEEEkeywords}
beyond 5G, deep learning, error performance, grant-free communication, multi-user detection, uplink NOMA.   
\end{IEEEkeywords}

% For peer review papers, you can put extra information on the cover
% page as needed:
% \ifCLASSOPTIONpeerreview
% \begin{center} \bfseries EDICS Category: 3-BBND \end{center}
% \fi
%\IEEEpeerreviewmaketitle
% For peerreview papers, this IEEEtran command inserts a page break and
% creates the second title. It will be ignored for other modes.
\section{Introduction}
Non-orthogonal multiple access (NOMA) scheme has a great potential for massive machine-type communication (mMTC) framework since it allows multiple users/devices to share the same resource block (RB).
However, especially in Internet of Things (IoT) networks, the data streams are generally sparse and low-rate. Thus, allocating an RB for an IoT device is not efficient owing to the limited RBs and thousands of IoT devices to be served by the same access point (AP)/base-station (BS). Besides, to grant an RB for an IoT device costs a signaling overhead. Furthermore, this signaling overhead should always be repeated since the number of active IoT devices changes from time to time due to the sparse communication, hence; this revokes the spectral efficiency provided by the NOMA. To resolve this, the grant-free NOMA schemes in uplink IoT networks have attracted great recent attention where the IoT devices transmit their low-rate data on the same RB (i.e., NOMA) without permission request \cite{Shahab2019}. 

The grant-free NOMA has been analyzed in terms of outage probability and capacity for finite and infinite alphabets \cite{Ding2018,Abbas2019a,Zhang2020}.However, the major drawback of the NOMA is its error performance due to the inter user interference (IUI). Indeed, this poor performance in uplink NOMA gets worse and an error floor occurs even in two-user networks \cite{Kara2019}. Nevertheless, the error performance of the grant-free NOMA has not been studied well although it has been widely analyzed in terms of information-theoretic perspectives. To the best of the authors' knowledge, the bit error rate (BER) performance of the grant-free NOMA schemes has not been evaluated yet.

On the other hand, the machine learning algorithms have gained great credibility for the last few years \cite{Ye2018} and their potentials for wireless communications have been admitted by the community \cite{Zappone2019}. Therefore, the machine learning algorithms, particularly deep learning (DL), have been proposed in NOMA-involved systems for modulation, constellation design and resource allocation \cite{Gui2018,Zhang2020a,Miao2020}. Besides, the detector designs\footnote{In the literature, code-domain NOMA has also been considered for the grant-free communication. Hence, the multi-user detection and end-to-end optimization of the code-domain NOMA empowered by DL have been investigated in \cite{DeepNOMA} where multiple RBs are used with preambles rather than one RB as being in power domain-NOMA. Thus, code-domain NOMA is beyond the scope of this paper and NOMA refers to power-domain NOMA.} via DL instead of conventional detectors have also been investigated for basic downlink \cite{Zhang2018b,Lin2019,Emir2019,Kang2020} and uplink \cite{Thompson2019,Emir} NOMA schemes. However, none of the previous uplink designs \cite{Thompson2019,Emir} considers a grant-free access where the number of devices is fixed to only two, which is very low for the IoT networks. Furthermore, in \cite{Thompson2019}, three frames are required to detect one frame data to acquire the channel information of the devices. This causes a spectral efficiency decay by $66.67\%$ and weakens the advantage introduced by the NOMA. To the best of the authors' knowledge, there is no study in the literature which provides a DL-based multi-user detection for uplink NOMA nor a spectral efficient practical design for a grant-free NOMA even in two-user network.

Motivated by the above discussions, we propose a DL-aided multi-user detection for grant-free IoT networks. The main contributions of this letter are summarized as follows. 
\begin{itemize}
    \item We propose a DL-aided multi-user detection (DeepMuD) which is able to detect symbols of an arbitrary number of devices in grant-free communication. The DeepMuD has been trained offline and implemented as an online detector in the grant-free IoT NOMA networks.
    \item With the proposed DeepMuD, the error performance of the uplink NOMA has been improved significantly. The same error performance with the conventional detectors for arbitrary number of users has been achieved with at least $10$ dB less power consumption which is very promising for energy-limited IoT devices.
    \item The proposed DeepMuD has a flexible detection structure where less than or equal to the number of devices in the training can be detected as online without performance degradation. Thus, the DeepMuD enables the grant-free communication without introducing a signaling overhead.
    \item In the proposed DeepMuD, the perfect CSI is also not required. The DeepMuD performs a multi-user detection based on pilot signals. The used pilot-to-frame ratio is also very low compared to the existing works. Thus, the advantage of the NOMA is ensured. 
\end{itemize}
The rest of the letter is organized as follows. In Section II, the uplink NOMA scheme is introduced (with the benchmark detector) and the signal design for the grant-free uplink NOMA is given. In Section III, the proposed DeepMuD is given in detail. Then, in Section IV, the error performance comparisons are presented between the proposed DeepMuD and the existing detectors. The results are discussed in detail. Finally, Section V concludes the letter. 
\section{System Model and Signal Design for Grant-Free NOMA}
\begin{figure}
		\centering
    \includegraphics[width=7cm, height=3.5cm]{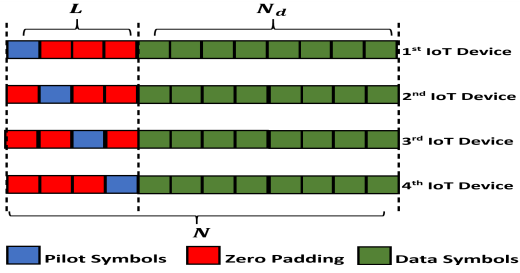}
    \caption{The frame design of IoT devices in grant-free uplink NOMA for $L=4$  $N_d=8$, $N=L+N_d=12$.}
    \label{EH_protocols}
\end{figure}
We consider an uplink scheme, where an AP and $L$ IoT devices (e.g., sensors within a factory or agriculture area) are located. We assume that all nodes are equipped with single antenna and the channel fading between the $l$th (i.e., $l=1,2,\dots,L$) device and the AP experiences a block-flat fading Rayleigh channel. Since a grant-free communication is considered, the active devices, which have data to convey, transmit their data to the AP on the same RB simultaneously. Hence, the received signal at the AP is given as
\begin{equation}
    y(t)=\sum_{i=1}^{K}\sqrt{P_i}x_i(t)h_i(t)+n(t), \quad K\leq L
\end{equation}
where $K$ is the number of active devices. $P_i$, $x_i(t)$ and $h_i(t)$ are the transmit power, transmitted symbol and channel coefficient of the $i$th active device, respectively. $n(t)$ is the additive white Gaussian noise (AWGN) which follows $CN(0, \sfrac{N_0}{2})$. 
\subsection{Successive Interference Canceler-based Detector (SICD) (Benchmark)}
In the SICD, the AP detects IoT devices' symbols in a successive manner, where the symbols of the IoT device with the best channel condition are detected firstly. Then, these detected symbols are subtracted from the received signal and the symbols of the IoT device with the second-best channel condition are detected. Thereafter, these secondly detected symbols are subtracted from the remaining received signal and this goes on until to the last device. Therefore, the detection process is given as
\begin{equation}
\begin{split}
    &\hat{x}_i(t)=\argmin_{j}{\left|y^{SIC,i}(t)-\sqrt{P_i}\hat{h}_i(t)x_{i,j}\right|^2},  \\
&\text{where} \quad j=1,2,\dots,M_i,\\
&y^{SIC,i}(t)=y^{SIC,i-1}(t)-\sqrt{P_{i-1}}\hat{h}_{i-1}(t)\hat{x}_{i-1}, \\
&\text{where} \quad  y^{SIC,0}(t)\triangleq y(t), \ \text{and}\\
&  \hat{x}_1=\argmin_{j}{\left|y(t)-\sqrt{P_1}\hat{h}_1(t)x_{1,j}\right|^2}, j=1,2,\dots,M_i, 
\end{split}
\end{equation}
where $\hat{x}_i$ is the detected/estimated symbol of the $i$th device. $x_{i,j}$ is the $j$th constellation point in $M_i$-ary modulation order of the $i$th device. $\hat{h}_i$ is the estimated (imperfect and/or perfect) channel coefficient of the $i$th device at the AP, thus; an extra channel estimation algorithm (e.g., LS, MMSE) should be implemented. As seen in (2), in the SICD, an additional latency is introduced since $K$ times maximum likelihood detection and $K-1$ times subtraction should be implemented in order.
\subsection{Frame/Signal Design with Pilot Insertions}
As explained above, in the SICD, an additional channel estimation algorithm is required for the CSI knowledge (since the SICD requires the channel order to perform detection). On the other hand, thanks to their high capacity in correlating, the DL-based detectors can perform well in joint channel estimation and symbol detection. However, the pilot insertion should be handled carefully not to cause an erroneous correlation, particularly in NOMA, since the IoT devices will be served simultaneously. To this end, we design frames\footnote{During one frame, the channel coefficients are assumed to be constant (e.g., block fading), which is quite reasonable since the IoT communication is generally performed in short-range communication with low-rate/frame size.} with pilot signals for each IoT device. In the frame designs of $L$ devices, we insert pilot signals and zero padding into the data frames. Hence, the total frame representations for the first and second (due to space limitations) IoT devices are given as
\begin{equation*}
\begin{split}
   &\mathbf{x}_1=\overbrace{\left[x_1^p(t),\underbrace{0,...,0,}_{L-1}\underbrace{x_{1,d}(t+L),...,x_{1,d}(t+N-1)}_{N_d} \right]}^N,\\
     &\mathbf{x}_2=\overbrace{\left[0,x_2^p(t+1),\underbrace{0,...,0,}_{L-2}\underbrace{x_{2,d}(t+L),...,x_{2,d}(t+N-1)}_{N_d}\right]}^N, 
\end{split}\tag3
\end{equation*}
where $x_i^p$ and $x_{i,d}$ denote the pilot signal and data symbol of the $i$th IoT device, respectively. The zero padding (e.g., $0,...,0$) is inserted since during these zero padding of the $i$th IoT device, the pilot signals for other devices (i.e., $x_{j}^p, \ j=1,2,...L, \ j\neq i$) are transferred to the AP. Hence, none of the pilot signals is overlapped so that the DeepMuD is able to correlate them with data symbols correctly. The overall frame design for $L=4$ is given in Fig. 1. Hereby, we would like to note that the pilot plus zero padding insertions into the frame could be interleaved (e.g., coordinated random interleaving). Nevertheless, we placed them in the beginning of the frame for the representation simplicity. Since the transmission is assumed to be frame by frame, according to (1), the received frame turns out to be\stepcounter{equation}
\begin{equation}
    \mathbf{y}=\sum_{i=1}^K\sqrt{P_i}h_i\mathbf{x}_i+\mathbf{n},
\end{equation}
where $\mathbf{y}=[y(t),y(t+1),...y(t+N)]$ and $\mathbf{n}=[n(t),n(t+1),...n(t+N)]$ are defined.

In (3) and (4), $N_d$ and $N$ are the data symbols' length and the total frame size, respectively. Therefore, during an $N$ frame size, $N_d$ symbols are transmitted for each IoT devices whereas $N-N_d=L$ symbols are used for non-data information. Thus, the \textit{ensured} ergodic capacity is given as
\begin{equation}
    C=\delta B\sum_{i=1}^K\log_2{M_i}(1-P_i(e)),
\end{equation}
where $\delta=\sfrac{N_d}{N}$ is the data-to-frame size ratio and $B$ is the bandwidth. $P_i(e)$ is the error probability of the $i$th IoT device. We introduce $P_i(e)$ in  capacity definition since the erroneous-detected symbols are not meaningful, thus we only consider correct-detected symbols such that calling \textit{ensured} capacity. 
\section{Proposed DeepMuD}
\begin{figure*}
		\centering
    \includegraphics[width=16cm, height=3.2cm]{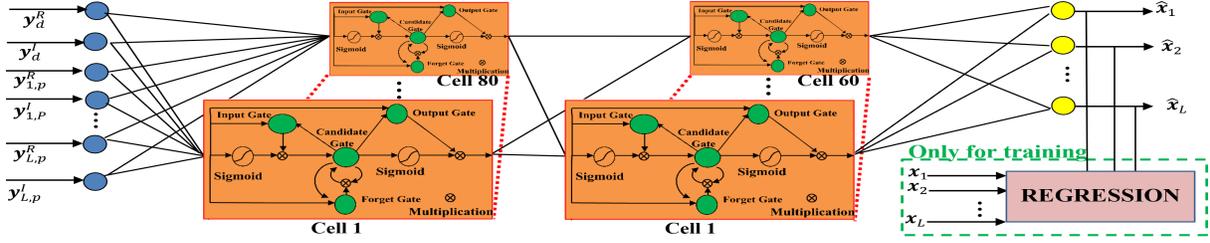}
    \caption{The proposed DeepMuD network model.}
    \label{lstm}
\end{figure*}
In the proposed DeepMuD, we use a model-driven DL to detect symbols simultaneously. The proposed model is based on a Long Short-Term Memory (LSTM) network. The LSTM is chosen since the LSTM networks have good performances in predicting frame size (time series) data where within the frame, the data is correlated \cite{Gui2018}. As explained above, in the considered model, we assume that the channel coefficients do not change within a frame. Therefore, the symbols are exposed the same channel effect and have correlation. The proposed model has 4 layers. The first layer is the input layer where the inputs are transferred to the further layers with the trained-weight coefficients. In the next two layers, we use LSTM layers which have $80$ and $60$ LSTM cells,\footnote{There is no theoretical way to find/select the optimum number of layers or cells in each layer \cite{Hochreiter1997}. Thus, as being in all DL-aided communications applications \cite{DeepNOMA,Gui2018,Emir2019,Emir,Zappone2019,Thompson2019,Zhang2018b,Miao2020,Lin2019,Kang2020,Ye2018,Zhang2020a}, these parameters are empirically determined, such that increasing the sizes do not provide a noteworthy gain in learning performance and the network performance converges.} respectively. The last layer is a fully-connected layer which produces the estimated symbols for each IoT device. Hence, it has $L$ neurons to transfer the estimated symbols to the outputs. Besides, we use a regression layer as a feedback in the training. The proposed model is presented in Fig. 2.  

\subsection{Dataset Generation}
Since the DL models can not perform with complex data, we split the received data into two parts as real and imaginary parts. Besides, in LSTM networks, the frame sizes of the input and output should be the same. In the considered system, we seek detecting data symbols with $N_d$ length, thus; the output frame size is $N_d$. Based on these, we reform the received signal and these reformed-inputs ($2$ inputs (real and imaginary) for received data symbols and $2L$ inputs (real and imaginary) for pilot responses of each device) vectors are given by 
%\begin{figure*}[t]
  \begin{equation*}
\begin{split}
  &\mathbf{y}_{i,p}^R=\mathpzc{Re}\{\overbrace{\left[y(t+i-1),y(t+i-1),...,y(t+i-1)\right]}^{N_d}\}, \\
    &\mathbf{y}_{i,p}^I=\mathpzc{Im}\{\overbrace{\left[y(t+i-1),y(t+i-1),...,y(t+i-1)\right]}^{N_d}\}, \\
    &\mathbf{y}_d^R=\mathpzc{Re}\{\overbrace{\left[y(t+L),y(t+L+1),...\dots, y(t+N-1)\right]}^{N_d}\}, \\
     &\mathbf{y}_d^I=\mathpzc{Im}\{\overbrace{\left[y(t+L),y(t+L+1),...\dots, y(t+N-1)\right]}^{N_d}\}.  
\end{split}\tag6
\end{equation*}\stepcounter{equation}
% \hrulefill
% \end{figure*}
In (6), we extend the pilot inputs (i.e., $\mathbf{y}_{i,p}^R$ and $\mathbf{y}_{i,p}^I$) to have $N_d$ length by adding copies of the received pilot responses.

The dataset generation is given in Algorithm 1, where $S$ is the sample size for each SNR value. Besides, to cover the effects of different SNR values, in the dataset, we define $\mathbf{SNR}=[0:5:30]$ dB. The outputs of Algorithm 1 are the $\mathbf{X}_{i,d}$,$\mathbf{Y}_{i,p}^R$,$\mathbf{Y}_{i,p}^I$,$\mathbf{Y}_d^R$,$\mathbf{Y}_d^I$ matrices and each rows of them are equal to the obtained vectors at Step 4 and Step 6 of Algorithm 1 for each iteration.  
\begin{algorithm}[ht]
\caption{Dataset Generation for Training DeepMuD}\label{dataset}
\begin{algorithmic}[1]
\State \textbf {Inputs} ($N,L,S, \mathbf{SNR}$)
\State \textbf {for} {each SNR value in $\textbf{SNR}$}
\State \textbf {for} {\textbf {$s=1:S$} }
\State Generate random $N_d\log_2M_i$ bits for $L$ IoT devices and maps them to $\mathbf{x}_{i,d}$ by $M_i$-ary modulation. Then according to (3) and Fig. 1, design $\mathbf{x}_i$ frame for each device.   
\State Generate random Rayleigh channel coefficients ($h_i$) for each device and generate random AWGN vector $\mathbf{n}$
\State According to (4), calculate the $\mathbf{y}$ and based on this $\mathbf{y}$, obtain the $\mathbf{y}_{i,p}^R$, $\mathbf{y}_{i,p}^I$,$\mathbf{y}_d^R$ and $\mathbf{y}_d^I$ by using (6).
%\State \textbf {end}
%\State \textbf {end}
\State \textbf {Outputs}  $\mathbf{X}_{i,d}$,$\mathbf{Y}_{i,p}^R$,$\mathbf{Y}_{i,p}^I$,$\mathbf{Y}_d^R$,$\mathbf{Y}_d^I$
\end{algorithmic}
\end{algorithm}
\subsection{Model Training}
After obtaining the training dataset, the proposed DeepMuD is trained according to DL parameters. In the training process, the outputs of the DeepMuD are expected to be equal to $\mathbf{x}_{i,d}$ where $ i=1,2,...,L$. Therefore, the training optimization problem is given by half mean square error as
\begin{equation}
   \{\mathcal{P}_1\}=\min \left\{\frac{1}{2L}\sum_{i=1}^L\left||\mathbf{x}_i-\mathbf{\hat{x}}_i|\right|^2\right\}.
\end{equation}

In addition, by considering the user fairness of IoT devices and not to cause severe error performance for any of the IoT devices, we define optimization problem in terms of error performance as
\begin{equation}
    \{\mathcal{P}_2\}=\min\left\{\max\{P_i(e)\}\right\}, \quad i=1,2,...,L.
\end{equation}

To this end, we initialize DL parameters and train the network according to $\mathcal{P}_1$ in (7). Then, the trained network is implemented as a detector in the simulations and the BER performances of each device are obtained. According to this BER performances, we update DL parameters and retrain the network unless the $\mathcal{P}_2$ in (8) is satisfied. The training and parameter optimization are given in Algorithm 2. The training settings in Algorithm 1 and Algorithm 2 and the optimized parameters are given in Table I. 
\begin{algorithm}[ht]
\caption{Training and Optimization of the DeepMuD}\label{traint}
\begin{algorithmic}[1]
\State \textbf {Inputs} ($\mathbf{X}_{i,d}$,$\mathbf{Y}_{i,p}^R$,$\mathbf{Y}_{i,p}^I$,$\mathbf{Y}_d^R$,$\mathbf{Y}_d^I$)
\State \textbf {Initialize DL Parameters} {(mini batch size, learning rate, maximum epoch)}
\State Train the network with the DL parameters according to $\mathcal{P}_1$ in (7)
\State  \textbf {do} Update DL training parameters and go to Step 3
\State \textbf {while} {Unless the $\mathcal{P}_2$ in (8) is satisfied }
\State \textbf {Outputs} {The DeepMuD network, optimized DL parameters}
\end{algorithmic}
\end{algorithm}
\begin{table}
\centering
\caption{Training Settings and Optimized Parameters}
\begin{tabular}{c||c}
\hline
Parameter &Value \\ \hline \hline
Programming & MATLAB \\ \hline
$\mathbf{SNR}$ (dB)& $[0:5:30]$ \\ \hline
Modulation & BPSK \\ \hline
Fading and Channel & Rayleigh+AWGN \\ \hline
IoT devices in a RB ($L$) & $4, \ 6$ \\ \hline
%Active IoT devices ($K$) & $2, \ 4$ for $L=4$ and $4, \ 6$ for $L=6$ \\ \hline
Frame Size ($N$) & $16, \ 64$ \\ \hline
%Data to Frame Size Ratio ($\delta$) & $\sfrac{1}{2}$, $\sfrac{2}{3}$ \\ \hline
Number of Samples ($S$) per scenario & $10^5$ \\ \hline
Optimizer& ADAM\\ \hline
Learning Rate & 0.001 \\ \hline
Mini Batch Size & 1000\\ \hline
Maximum Epoch &20\\ \hline
\end{tabular}
\end{table}
\vspace{-1\baselineskip}
\subsection{Complexity}
We focus on the online implementation complexity (i.e., feed-forward calculations) for the DeepMuD. The computational complexity for an LSTM-based network is given by $\mathcal{O}(W)$ where $W$ is the weight size in the hidden layer \cite{Hochreiter1997}. We use two hidden layers, hence the computational complexity of the DeepMuD is obtained as $\mathcal{O}(W_1W_2)=\mathcal{O}(80\times60)$. On the other hand, the computational complexity of the SICD is given as $\mathcal{O}(4ML+2(L-1))$. Besides, in the SICD, an additional complexity will be introduced due to the channel estimation algorithm which is not required in the DeepMuD. It is seen that the complexity of the SICD is increased by the modulation order and the number of IoT devices whereas the proposed DeepMuD has the same complexity regardless of these. On the other hand, with the lower $M$ and $L$, the DeepMuD may have a bit higher complexity. Nevertheless, we should note that the detection is performed at the AP and the AP has high computational capacity. Besides, considering the performance gain in the DeepMuD (see the next section), this negligible complexity increase is affordable.
\vspace{-0.5\baselineskip}
\section{Numerical Results}
In this section, we implement the offline-trained DeepMuD as an online detector and perform experiments on synthetic dataset. The link-level Monte-Carlo simulations\footnote{In simulations, all IoT devices have equal transmit power (i.e., $\rho=\rho_i=\sfrac{P_i}{N_0}, \forall i$) and the channel conditions are $E[|h_{i}|^2]=E[|h_{i+1}|^2]+3$ dB and $E[|h_{K}|^2]=0$ dB.} are conducted to validate the performance comparisons.

In Fig. 3, we present performance comparisons between DeepMuD and SICD for $L=4,6$ where BER and capacity comparisons are given in Fig. 3.a, 3.b and in Fig. 3.c, respectively. We assume that the number of active users are equal to the number of total users to be served (i.e., $ K=L$). Besides, in all subfigures, we present the results of DeepMuD for different frame sizes ($N$). As seen in Fig. 3.a and 3.b, regardless of the frame size ($N$) and/or number of IoT devices in a RB ($L$), the proposed DeepMuD always outperforms the SICD significantly even though perfect CSI is assumed in SICD whereas only pilot responses are available for the proposed DeepMuD. Hereby, it is worth noting that the SICD can not detect any symbols when the number of devices is higher than $2$. (In Fig. 4, the SICD works for only two-user networks. However, it still has an error floor.) On the other hand, once the DeepMuD results are compared, with the increase of frame size ($N$), the performance is improved in $K=L=4$ in Fig. 3.a since the correlation between data increases so that the network learns better. Nevertheless, due to the IUI, this gain is smaller when $K=L=6$ in Fig. 3.b.
\begin{figure*}
\centering
\subfloat[{BER, $K=L=4$}]{\includegraphics[width=6cm]{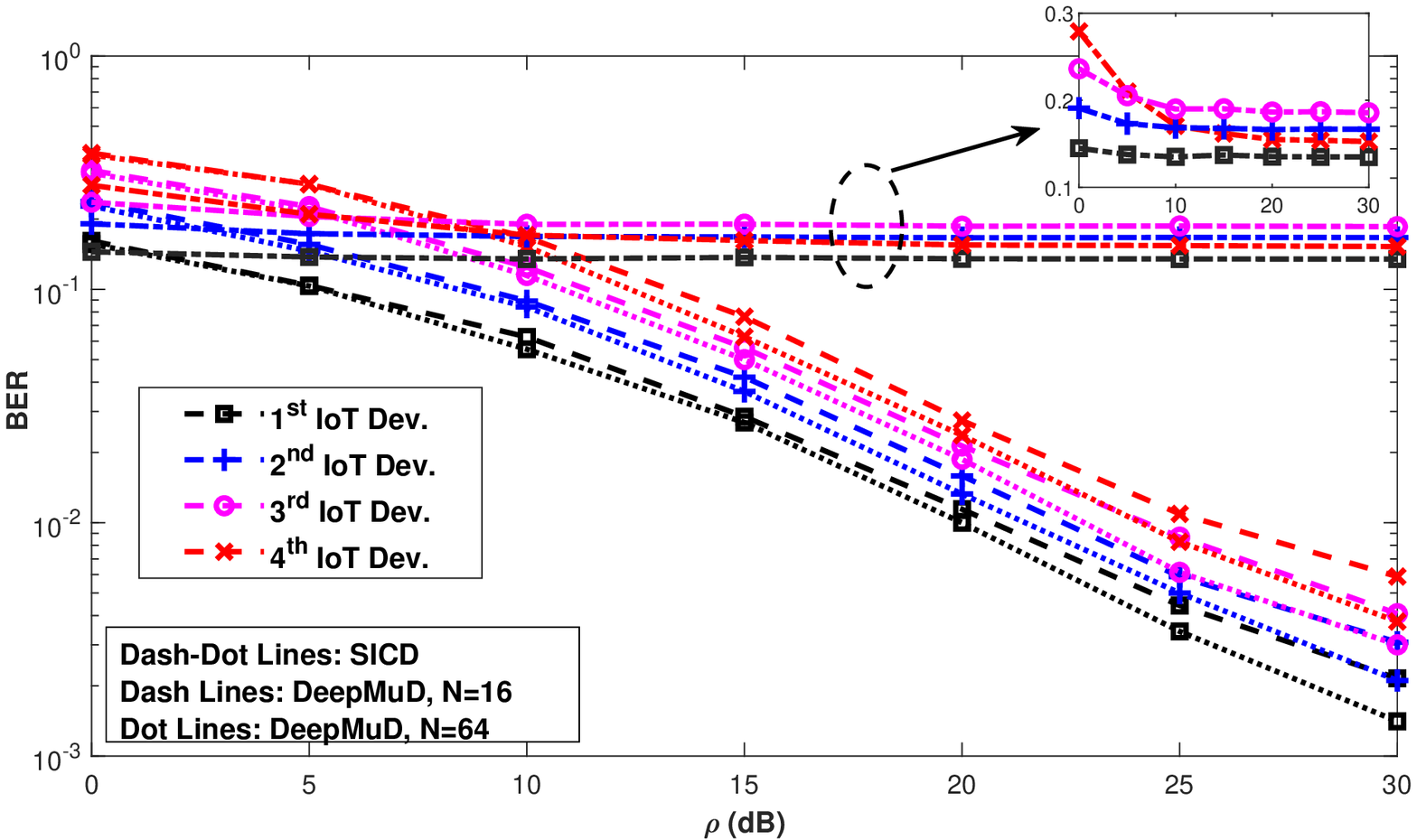}
\label{ber1}}
\subfloat[{BER, $K=L=6$}]{\includegraphics[width=6cm]{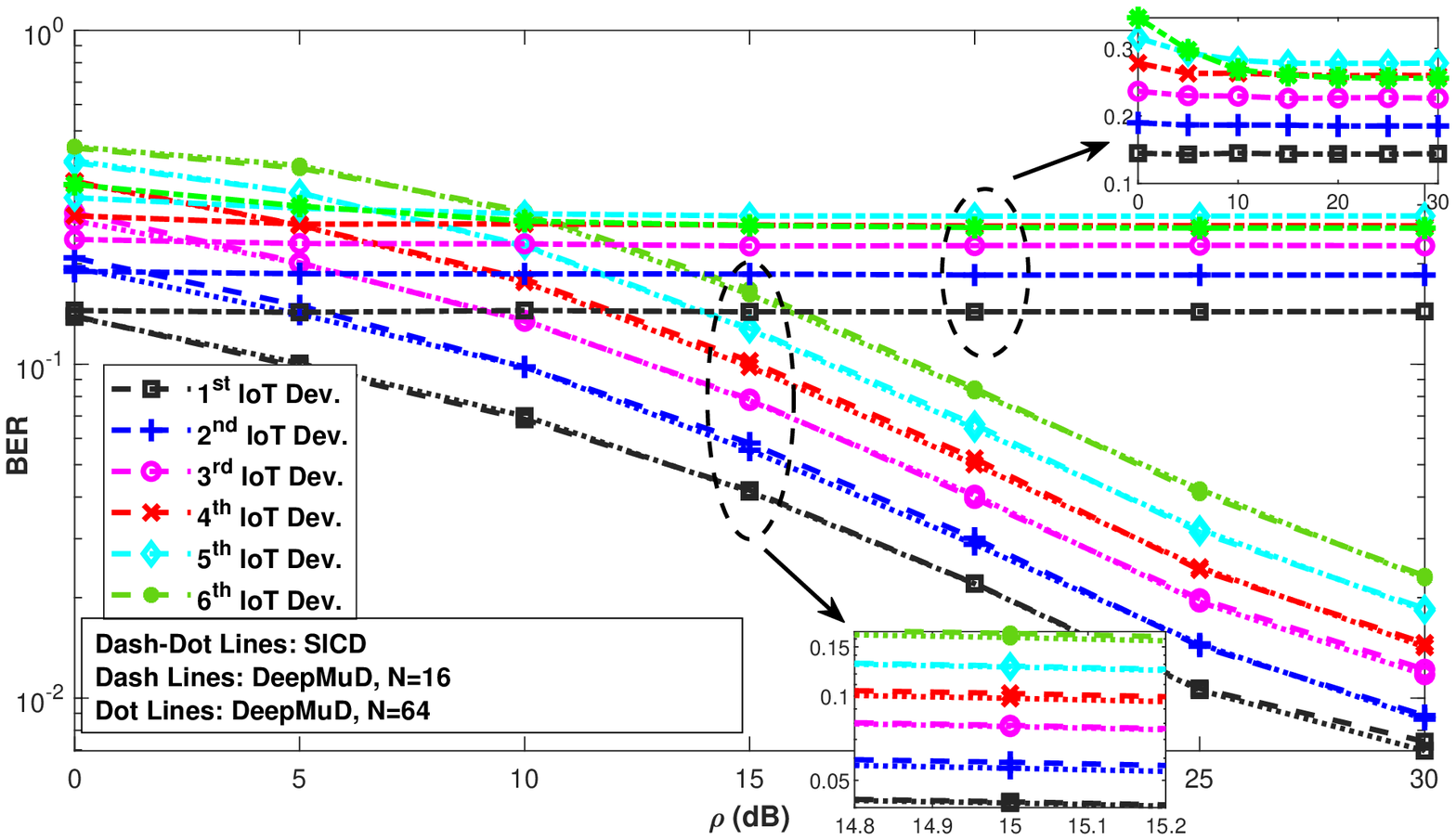}
\label{ber2}}
\subfloat[{Normalized Ensured Capacity, $K=L=4,6$}]{\includegraphics[width=6cm]{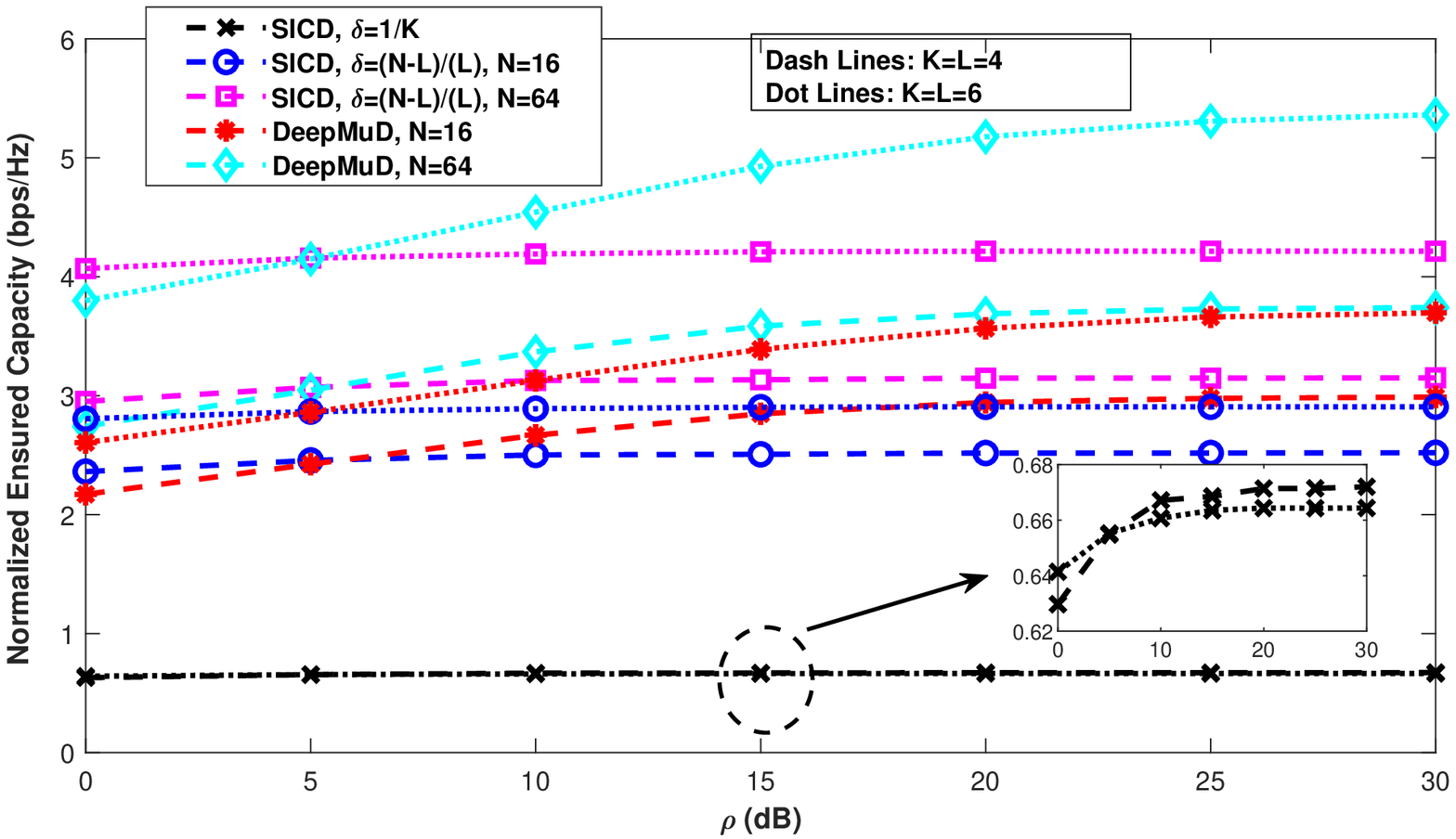}
\label{diversity}}
\caption{{Performance comparisons of DeepMuD and SICD when $K=L=4,6$}}
\label{vs_SNR}
\end{figure*}

Then, to present the \textit{ensured} capacity as defined in (5), we provide normalized (i.e., $B=$1Hz) \textit{ensured} capacity comparisons between the DeepMuD and SICD in Fig. 3.c. Since the SICD performs symbol-by-symbol detection, it requires $K$ time slots for the CSI knowledge to detect one symbol. Thus, $\delta$ in SICD is equal to $\sfrac{1}{(K+1)}$. Nevertheless, we also give SICD results for the same $\delta=\sfrac{N_d}{N}$ as in the proposed model. One can easily see that the proposed DeepMuD is superior to SICD also in terms of \textit{ensured} capacity where the SICD can never achieve the theoretical capacity. This is explained as follows. In uplink NOMA, it is theoretically possible to improve the capacity; however, without proper detection, the BER performance is severe. Therefore, the achievable capacity is actually non-detectable capacity. Besides, it is noteworthy that the perfect CSI is assumed in SICD and its performance will get worse when it is relaxed. On the other hand, the proposed DeepMuD provides an increased BER performance with low pilot-to-frame size ratio ($1-\delta$); hence, the achievable capacity is ensured.

Lastly, in order to reveal the flexibility of the proposed DeepMuD, in Fig. 4, we present BER performances for the  scenarios in which the number of active IoT devices are equal to $K=2$ and $K=4$ when the training is performed for $L=4$ and $L=6$, respectively. As seen in Fig. 4, the proposed DeepMuD has performed well even though the number of active devices are less than the training numbers and the DeepMuD outperforms the SICD remarkably. This proves the power of the DeepMuD for grant-free networks since the DeepMuD can detect symbols for arbitrary number of IoT devices once it is trained offline. Therefore, the IoT devices do not require a grant permission thus reducing the signaling overhead. Hereby, we note that the minimum performance gain provided by the DeepMuD is achieved when $K=2$ where the DeepMuD and SICD have the same BER performances with $\sim10$ dB less power consumption in DeepMuD. With the higher $K$, the DeepMuD provides much more performance gain and it can be up to $30$ dB in some scenarios.
\begin{figure}
\centering
\subfloat[$K=2$, $L=4$]{\includegraphics[width=4cm, height=3.5cm]{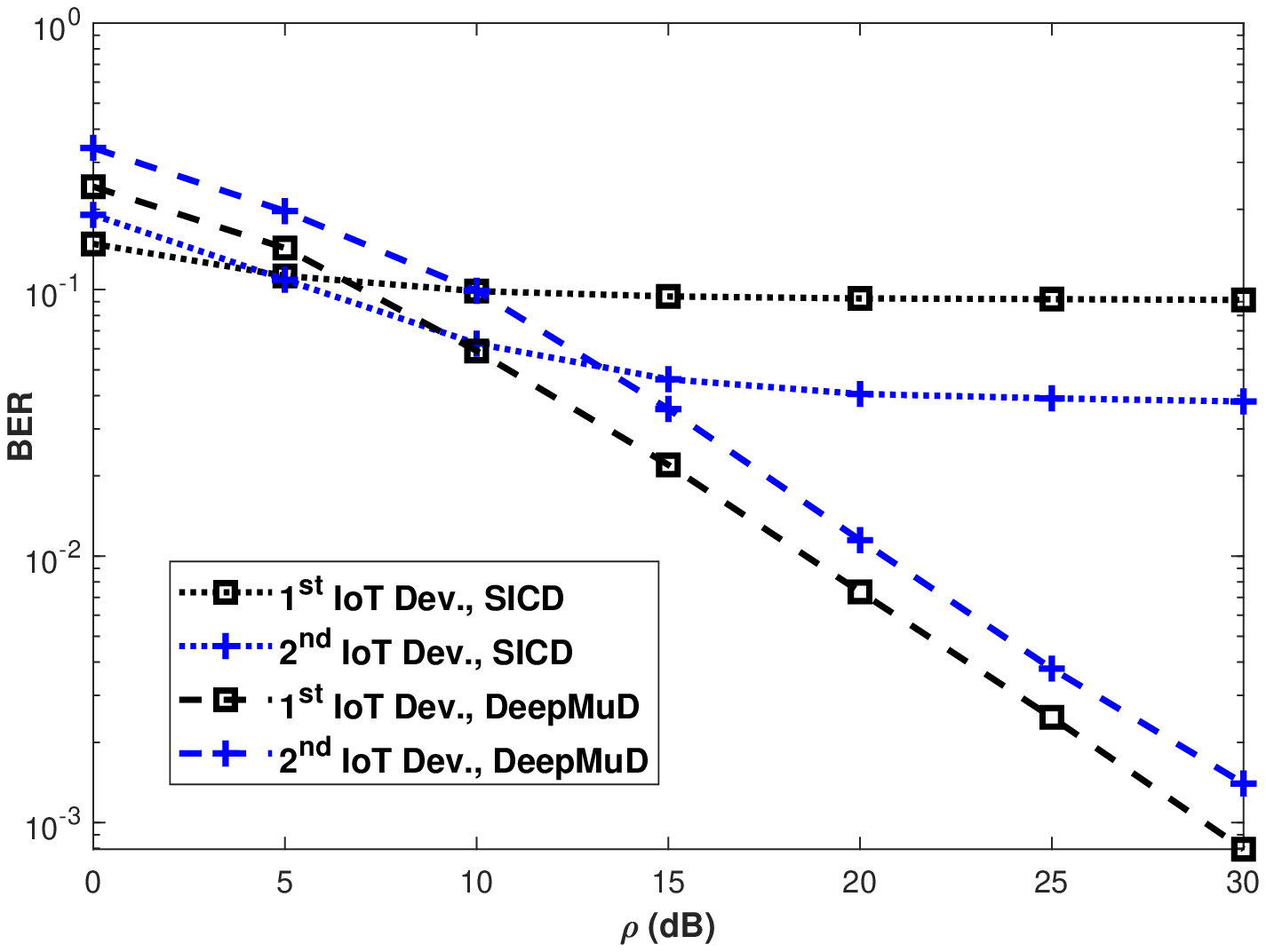}
\label{K_2_L_4}}
\subfloat[$K=4$, $L=6$]{\includegraphics[width=4cm, height=3.5cm]{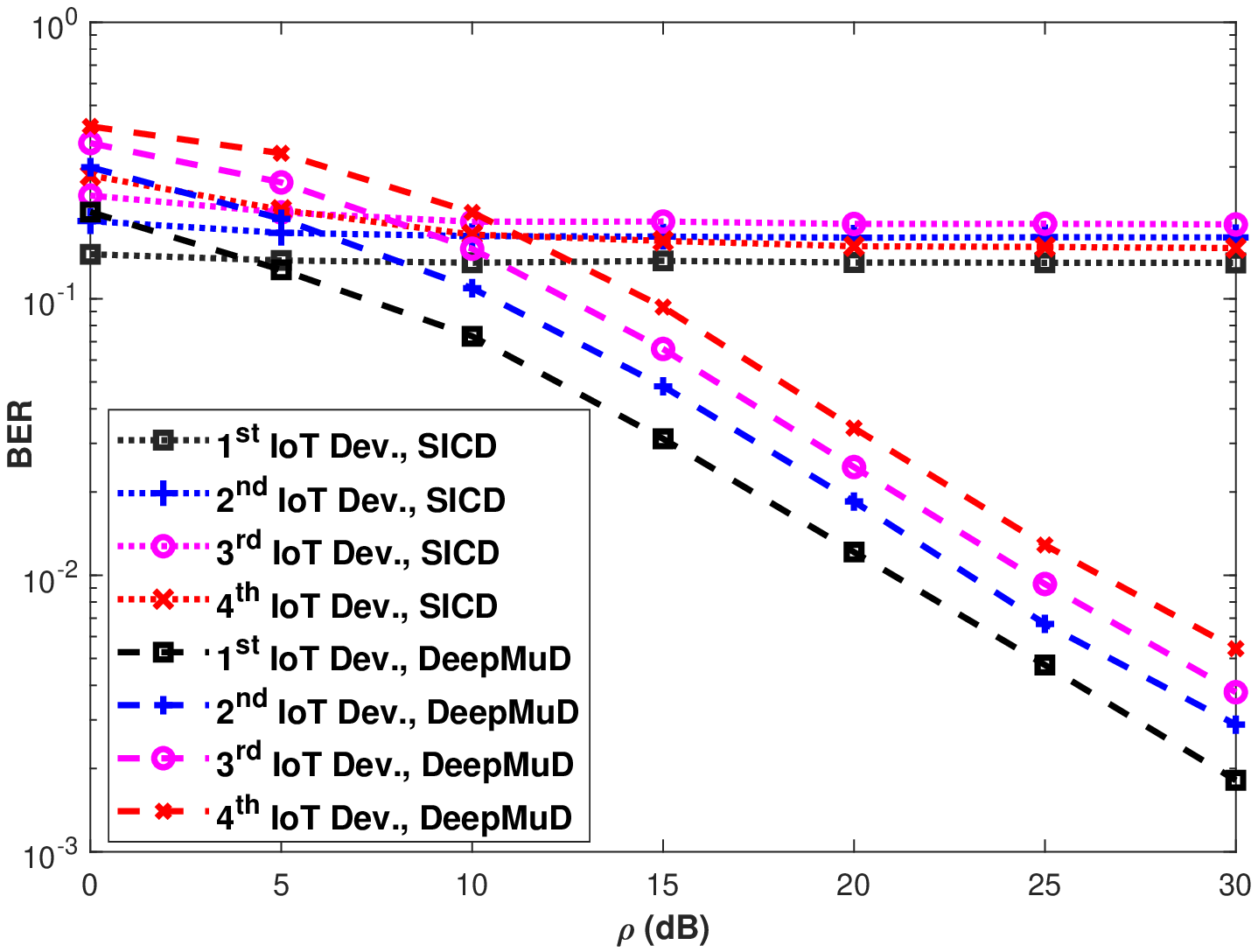}
\label{K_4_L_6}}
\caption{BER comparisons of DeepMuD and SICD.}
\label{flexible}
\end{figure}
\section{Conclusion}
In this paper, we propose a DL-aided multi-user detection (DeepMuD) in grant-free uplink IoT NOMA networks. The proposed DeepMuD has a good BER performance and outperforms existing multi-user detection schemes. Besides, with the proposed DeepMuD, the signal detection can be performed for arbitrary number of IoT devices (less than and/or equal to the number in training, i.e., $K\leq L$) which enables the grant-free access. Furthermore, the DeepMuD detects signal based on pilot responses, therefore; no additional channel estimation algorithm is needed. This reveals the power of DeepMuD in joint signal detection, and it can be applied in other promising physical layer schemes such as faster than Nyquist, index modulations, and large intelligent surfaces. These are seen as future research directions.
%\numberwithin{equation}{section}
%\makeatletter
%% "activate" the preparatory code, but for section-level headers only
%\newcommand{\section@cntformat}{Appendix \thesection:\ }
%\makeatother

% if have a single appendix:
%\appendix[Proof of the Zonklar Equations]
% or
%\appendix  % for no appendix heading
% do not use \section anymore after \appendix, only \section*
% is possibly needed
% use appendices with more than one appendix
% then use \section to start each appendix
% you must declare a \section before using any
% \subsection or using \label (\appendices by itself
% starts a section numbered zero.)
%
% you can choose not to have a title for an appendix
% if you want by leaving the argument blank
% use section* for acknowledgment
% Can use something like this to put references on a page
% by themselves when using endfloat and the captionsoff option.
\ifCLASSOPTIONcaptionsoff
  \newpage
\fi
% trigger a \newpage just before the given reference
% number - used to balance the columns on the last page
% adjust value as needed - may need to be readjusted if
% the document is modified later
%\IEEEtriggeratref{8}
% The "triggered" command can be changed if desired:
%\IEEEtriggercmd{\enlargethispage{-5in}}
% references section
% can use a bibliography generated by BibTeX as a .bbl file
% BibTeX documentation can be easily obtained at:
% http://mirror.ctan.org/biblio/bibtex/contrib/doc/
% The IEEEtran BibTeX style support page is at:
% http://www.michaelshell.org/tex/ieeetran/bibtex/
%\bibliographystyle{IEEEtran}
% argument is your BibTeX string definitions and bibliography database(s)
%\bibliography{IEEEabrv,../bib/paper}
%
% <OR> manually copy in the resultant .bbl file
% set second argument of \begin to the number of references
% (used to reserve space for the reference number labels box)
\vspace{-.7\baselineskip}
\bibliographystyle{IEEEtran}
\bibliography{kara_WCL2021_0187}
% biography section
%
% If you have an EPS/PDF photo (graphicx package needed) extra braces are
% needed around the contents of the optional argument to biography to prevent
% the LaTeX parser from getting confused when it sees the complicated
% \includegraphics command within an optional argument. (You could create
% your own custom macro containing the \includegraphics command to make things
% simpler here.)
%\begin{IEEEbiography}[{\includegraphics[width=1in,height=1.25in,clip,keepaspectratio]{mshell}}]{Michael Shell}
% or if you just want to reserve a space for a photo:
% if you will not have a photo at all:
% insert where needed to balance the two columns on the last page with
% biographies
%\newpage
% You can push biographies down or up by placing
% a \vfill before or after them. The appropriate
% use of \vfill depends on what kind of text is
% on the last page and whether or not the columns
% are being equalized.
%\vfill
% Can be used to pull up biographies so that the bottom of the last one
% is flush with the other column.
%\enlargethispage{-5in}
% that's all folks
\end{document}